\documentstyle[12pt,aasms4]{article} 
\tighten 
\newcommand{\be}{\begin{equation}} 
\newcommand{\ee}{\end{equation}}

\newcommand{\bea}{\begin{eqnarray}} 
\newcommand{\eea}{\end{eqnarray}} 
\newcommand{\eg}{e.g., } 
\newcommand{\ie}{i.e., } 
\newcommand{\OM}{\Omega_m} 
\newcommand{\OO}{\Omega_0}

\newcommand{\nn}{\nonumber\\} 
\newcommand{\OL}{\Omega_{\Lambda}} 
\newcommand{\bO}{b_{\Omega}}

\newcommand{\mz}{$m$-$z$\ } 

\begin{document}
\title{THE EFFECTS OF INHOMOGENEITIES ON EVALUATING \\ 
THE MASS PARAMETER $\OM$ AND THE COSMOLOGICAL CONSTANT $\Lambda$} 
\author{ R.  Kantowski \altaffilmark{1}}
\affil{ University of Oklahoma, Department of Physics and
Astronomy,\\ Norman, OK 73019, USA }
\altaffiltext{1}{kantowski@mail.nhn.ou.edu}
\authoremail{kantowski@mail.nhn.ou.edu}

\begin{abstract} \vskip .2 truein Analytic expressions for distance--redshift relations which have
been corrected for the effects of inhomogeneities in the Friedmann-Lema\^itre-Robertson-Walker
(FLRW) mass density are given in terms of Heun functions and are used to illustrate the significance of inhomogeneities on a
determination of the mass parameter $\OM$ and the cosmological constant $\Lambda$.  The values of
these parameters inferred from a given set of observations depend on the fractional amount of
matter in inhomogeneities and can significantly differ from those obtained by using the
standard magnitude-redshift ($m$-$z$) result for pure dust FLRW models.  As an example 
a determination of $\OM$ made by applying  the  homogeneous distance--redshift relation to 
SN 1997ap at $z=0.83$ could be as much as 50\% lower than its true value.
\end{abstract}


\section{INTRODUCTION} \label{sec-intro} 
When attempting to evaluate the mass parameter $\OM$
and/or the cosmological constant $\Lambda$, observations of quantities such as magnitude, angular
separation, and redshift are made on objects distant enough for curvature effects to be detected.
As an example, for Type Ia supernovae (SNe Ia) corrected magnitudes and redshifts ($m$-$z$) are
measured, plotted, and compared with theoretical $m(\OM,\Lambda; z)$ curves computed for the FLRW
models (\cite{PS1}, \cite{PS2}, \cite{GP}).  In spite of the fact that the FLRW models contain 
only homogeneously and isotropically
distributed perfect fluid gravity sources, one of these  models is assumed to represent the ``large
scale" geometry of the universe.  Relations like $m(\OM,\Lambda;z)$ are also commonly assumed to
be valid, on average.  This latter assumption may well be incorrect for some distant observations
including SNe Ia, but even if technically correct may not be useful in determining $\OM$ and
$\Lambda$.  In particular if the underlying mass density approximately follows luminous matter
(\ie associated with bounded galaxies) then effects of inhomogeneities on relations like
$m(\OM,\Lambda;z)$ must be taken into account.  The majority of currently observed SNe Ia are not
being seen through foreground galaxies and whether or not this is due entirely to selection
(rather than statistics) is not important.  If the objects observed do not have the average FLRW
mass density $\rho_0$ in their foregrounds then the FLRW \mz\ relation does not apply to them 
(see \cite{KR2}).
Ultimately some SNe Ia should exist behind foreground galaxies (\cite{RK}) and for these, \mz\ should be
computed using the lensing formulas.  These formulas (\cite{BR} and \cite{CJ}) contain 
source-observer,
deflector-observer, and source-deflector distances, respectively $D_s, D_d$, and $D_{ds}$, all of
which depend on the mass density in the observing beam, \underline{excluding} the deflector.
These distances will not be given by the standard FLRW result if the observing beam contains less
than the average FLRW mass density.

In \S \ref{sec-optics} the average area-redshift equation (\ref{Af}) for a light beam traveling
through a FLRW Swiss cheese universe is given and its solution is related to the luminosity
distance--redshift relation $D_{\ell}(z)$.  In \S \ref{sec-B=0} the solution of this equation is
given for the case where gravitational lensing can be neglected.  The new result of this paper,
$D_{\ell}(z)$ without lensing for FLRW Swiss cheese can be found in equations (\ref{Dell0}) and
(\ref{Dellinfty}), and for the special case $\OO=1$ in equations (\ref{Dell01}) and
(\ref{Dellinfty1}) of Appendix A.  In \S \ref{sec-mzplots}  
numerous \mz plots are given to illustrate the
effects of inhomogeneities and some conclusions are drawn.  
It is argued that if homogeneities are not taken into
account when attempting to determine $\Omega_m$ and $\Lambda$, errors as large as 50\% could be
made.  Even though the $D_{\ell}(z)$ given here has been derived using the exact Swiss cheese
cosmologies, the result are  valid for observations in essentially any perturbed 
pressure-free FLRW
models in which lensing can be neglected.

Inhomogeneous models of the Swiss cheese type and their associated optical equations  discussed
here are often mistakenly attributed to Dyer and Roeder (see Appendix B). 

Appendix C contains some useful simplifications for evaluating the real-valued Heun functions
needed in the analytic \mz relations given here.  Appendix C also contains six useful lines of 
Mathematica code which numerically evaluates and plot these same \mz relations.

\section{SWISS CHEESE OPTICS} \label{sec-optics}

Some years ago the author (\cite{KR}), used the ``Swiss cheese"
cosmologies to study the effects of local inhomogeneities in the FRW mass density on the
propagation of light through an otherwise homogeneous and isotropic universe.  That analytic work
was undertaken because prior results computed in perturbed FRW models were suspect.
In particular, results were inconsistent with weak lensing results, \eg on average, the
luminosity of a distant object was not given by the FRW result (\cite{BB}).  Numerous sources of error were
suggested but particularly mistakes inherent in using perturbative gravity were suggested.  For
example the FRW relations between radius, redshift, and affine parameters were 
(and still are
for approximate GR solutions)
assumed valid in the presence of mass perturbations. At that time \cite{KR} put to rest any question of the possible
existence of an effect on the mean luminosity; theoretically it could exist!  Because the Swiss
cheese models are exact solutions to the Einstein equations the accuracy of the FRW relations
between radius, redshift, and affine parameters could be directly established.  Today Swiss
cheese itself is under attack as the source of the `erroneous' prediction (see \cite{FJ} and
\cite{WJ}, and additionally see the related work of \cite{PP}).  The intent of this particular paper is not to defend Swiss cheese
\underbar{predictions} against these attacks, that can be done elsewhere.  The reader should not
dismiss distance--redshift predictions made by Swiss cheese models because of their 
non-physical distribution of matter, \eg cheese and holes.  Because these models contain
the only two types of gravitational curvature (Ricci and Weyl) that affect optical
observations, and because they are fairly flexible in including density perturbations, they
\underline{should} adequately describe optical observations at least as far as $z=1$ 
(a distance to which galaxies and other inhomogeneities are thought to have undergone 
only minor changes).  The conjectured extension of the validity of
the optical equation (\ref{Af})  used here beyond the Swiss cheese models has been made frequently
since it was first derived and is argued by \cite{SEF} in their Sections 4.5.2 and 4.5.3.  

The purpose of this paper is
to extend analytic results for distance--redshift relations in inhomogeneous FRW models to FLRW,
\ie to include the cosmological constant (see \cite{KVB} and
\cite{SS}).  \cite{KR} derived a 2nd order intergal-differential equation [see (43a) of that paper or (1) of \cite{KVB}
as well as the equivalent 3rd order differential equation, (43b) of that paper or (6) of \cite{KVB}] for the
average cross-sectional area $A$ of a beam of light starting from a distant source and
propagating through a $\Lambda=0$ Swiss cheese universe (see Figure 1).  The solution of this
equation, with appropriate boundary conditions, gives all average quantities relating to
distance--redshift.  This equation and its derivation were easily extended to include a
cosmological constant by \cite{DC3}.  
However, as will be seen below,
extension of the equation's analytic solutions of \cite{KVB} and \cite{SS} is somewhat involved and the special
functions required are much less familiar to the math/physics community.  Weinberg sign
conventions will be again used (\cite{MTW}).

As a light beam from a distant SN Ia propagates through the universe 
(Figure 1) the cheese of the model produces the same focusing effect 
as does the  transparent 
material actually appearing within the beam. The holes in the cheese 
with their condensed central masses
reproduce the optical effects of the remaining Friedmann matter that has been condensed 
into clumps, \eg galaxies.
The extended equation for the average area $A$ traversing the universe, that is randomly focused by numerous 
clouds of transparent matter and lensed by numerous clumps is:  
\be
{\sqrt{A}^{''}\over \sqrt{A}} +
{\langle\xi^2\rangle\over A^2} = -{3\over 2}\ {\rho_D\over \rho_0}\ \Omega_m\ (1+z)^5\ ,
\label{Af} 
\ee
where prime ($'$) is differentiation with respect to an affine parameter, 
\be
{'}\
\equiv -(1+z)^3 \sqrt{1+\OM z+\OL[(1+z)^{-2}-1]} \ {d \ \over d z}\ , \label{affine} 
\ee
and
$\langle\xi^2\rangle$ is the average of $(\sigma/A)^2$, the square of the wavefront's shear over
its area,\footnote[2]{The form of $\langle\xi^2\rangle$ depends on structure details of the clumps.
 What is given in (3) is for objects completely condensed into opaque spherical masses.
  This particular type of  Swiss cheese clumping is expected to produce maximum lensing.
  In the following sections we will be interested in observations where even maximum 
  lensing is negligible.} 
\be
\langle\xi^2\rangle = {15\over 2}\ {\rho_I\over \rho_0}\ B_0\ \Omega_m \int \
{A^2(1+ z)^6\over z'} d z\,.  \label{xi} 
\ee
In (\ref{Af}) $\rho_D$ ($D$ is for dust) is the
average mass density of all transparent material interior to light beams used to observe the given objects
and $\rho_I$ ($I$ is for inhomogeneous) is the average mass density of all types of clumpy
material systematically or statistically excluded from the light beams.
The average shear term in (\ref{Af}) comes from the Weyl (conformal) curvature tensor of inhomogeneous
material exterior to the beams. The  ${\rho_D/\rho_0}$ term comes 
from the Ricci tensor of transparent material within the beams.  
For the SNe Ia observations, $\rho_D$ would certainly include those ubiquitous 
low mass neutrinos (if they exist)
as well as other 
transparent material not confined to galaxies, while $\rho_I$ would contain all 
matter clumped with galaxies.
If there is no correlation of mass and light, deciding what goes in $\rho_D$ and what goes in 
$\rho_I$ is problematic, and the relative value becomes another unknown parameter of the theory.  
The current Friedmann
mass density is the total $\rho_0=\rho_D+\rho_I$ and the curvature parameter $\OO\equiv \Omega_m
+\Omega_{\Lambda}$ consists of a mass part and a cosmological constant part:  
\be
\Omega_m \equiv
{8\pi G \rho_0\over 3H_0^2} \hskip .25 in {\rm and } \hskip .25 in \Omega_{\Lambda} \equiv
{\Lambda c^2\over 3H_0^2}.  
\ee
Inclusion of the cosmological constant $\Lambda$ (using FLRW
rather than FRW) only modifies the functional relationship between redshift and affine parameter
(\ref{affine}), see \cite{DC3}.  The unitless gravitational lensing parameter $B_0$ is defined in equation
(A2) of \cite{KVB} and its effects on the solution of equation (\ref{Af}) are described in \cite{DC3}.  
 In this paper analytic solutions to equation (\ref{Af}) will be given for $B_0=0$, \ie for distance--redshift
 in any Swiss cheese model where $\langle\xi^2\rangle$ is neglected. In \cite{KVB} it was argued that even maximal lensing effects ($B_0\ne 0$) are not expected to be significant when
observing SNe Ia at $z\le 1$ and, as pointed out above, when lensing events do occur, the
observed magnitudes should be analyzed by using the lensing formula and not by incorporation into
the \mz\ relation. 
The resulting $B_0=0$ equation (\ref{Ab0}) represents the equation for the average 
area of a light beam only Ricci focused by part of the mass density, $\rho_D(\le\rho_0)$.
Such a light beam is not conformally lensed by inhomogeneities ($\rho_I=\rho_0-\rho_D$)
that remain exterior to the beam. 
If Weyl lensing is infrequent, 
 a distribution of areas will occur for which the $B_0=0$ 
equation gives the maximum value for the area (\ie the lower bound on the distribution of 
luminosities). 
In \cite{KR2} the resulting \mz is appropriately dubbed the `intergalactic' magnitude-redshift 
relation because it is \mz without galactic focusing. If significant galactic lensing is an unusual 
event as apparently is the case with 
SNe Ia beams passing exterior to galaxies, the `intergalactic' \mz approximates the  `mode' value 
(the most likely).\footnote[3]{This assertion is consistent with that part of
the numerical work of \cite{HD} which treated galaxies as condensed objects. 
When galaxies were modeled by 200 kpc isothermal spheres the `mode' moved 
towards the `mean' and away from
the minimum (intergalactic) value as expected of a more homogeneous model.} 
If galaxies are compact (20 kpc) the `intergalactic' \mz relation 
should be more useful in determining $\OM$ and $\OL$ 
from SNe Ia  observations than is the mean \mz relation (standard FLRW relation). 
If galaxies are more diffuse (200 kpc) exact modeling of the lensing galaxies will be important. 

To relate the differential equation (\ref{Af}) to observations, consider a source at redshift
$z_s$ radiating power ${\delta\cal P}$ into solid angle $\delta\Omega$.  The flux received by an
observer at $z=0$ in area $A|_0$ is given by ${\cal F} = {\delta\cal P}/ A|_0(1+z_s)^2$.  The two
factors of $(1+z_s)$ can be thought of as coming separately from the redshift of the observed
photons and their decreased rates of reception.  The definition of luminosity distance is
motivated by this result, \ie 
\be
D_{\ell}^2\equiv {A\big|_0 \over \delta\Omega}(1+z_s)^2.
\label{Dl} 
\ee
The observed area $A|_0$ is evaluated by integrating equation (\ref{Af}) from the
source $z=z_s$ to the observer $z=0$ with initial data which makes the wave front satisfy
Euclidean geometry when leaving the source (area=radius$^2\times$ solid angle):  
\bea
\sqrt{A}|_s&=&0,\nonumber\\ {d\sqrt{A\big|_s }\over dz}&=& -\sqrt{\delta\Omega} {c\over
H_s(1+z_s)}, \label{Aboundary} 
\eea 
where in FLRW the value of the Hubble parameter at $z_s$ is
related to the current value $H_0$ at $z=0$ by 
\be
H_s=H_0(1+z_s)\ \sqrt{1+\OM
z_s+\OL[(1+z_s)^{-2}-1]}.  \label{Hs} 
\ee
The series solution of equation (\ref{Af}), combined
with (\ref{Dl}) and (\ref{Aboundary}) is:
 \bea D_{\ell}(\OM,\OL,\nu,B_0;z)&=& \sqrt{{A\over
\delta\Omega}}\Bigg\vert_0(1+z) ={c\over H_0} \Biggl\{ z + {1\over2}\left[1+\OL-{1\over
2}\OM\right] z^2 \nonumber\\ &+&{1\over2} \left[ {1\over2}\OM \left(
{1\over2}\OM+{\nu(\nu+1)\over 6}-1 \right) - \OL\biggl(1+\OM-\OL\biggr) \right] z^3\nonumber\\
&+&{1\over8} \Biggl[ \OM\left( {1\over 8}\OM \biggl[10 - 2\nu(\nu+1)-5\OM\biggr]
-B_0{\nu(\nu+1)\over 6} \right)+\nonumber\\ &&\hskip .5 truein\OL \left( 5+{1\over
2}\biggl[5+\nu(\nu+1) \biggr]\OM +{15\over 4}\OM^2 +5 \OL^2-{5\over 2}\OL\biggl[4+3\OM\biggr]
\right) \Biggr] z^4\nonumber\\ &+&O[z^5] \Biggr\}, \label{Dlseries} \eea where the source
redshift $z_s$ has been simplified to $z$ and ${\rho_I/ \rho_0}$ has been replaced for later
convenience by a clumping parameter $\nu$, $0\le\nu\le 2$, 
\be
\nu\equiv
{\sqrt{1+24(\rho_I/\rho_0)}-1\over 2} \Rightarrow {\rho_I\over \rho_0}={\nu(\nu+1)\over 6}\ee
This series is useful for understanding the low-redshift sensitivity of $D_{\ell}$ to the various
parameters; \eg $\OM$ and $\OL$ appear in the $z^2$ term, the clumping parameter $\nu$ first
appears in the $z^3$ term, whereas the lensing parameter $B_0$ doesn't appear until the $z^4$
term.  Additionally, analytic results computed in the next section can be checked by comparison
with this series.
\section{THE ANALYTIC SOLUTION FOR \hbox{{\it D}$_{\ell}$}(${\lowercase {z}}$) 
WHEN LENSING CAN BE NEGLECTED} \label{sec-B=0}
In this section the general $B_0=0$ solution of (\ref{Af}) will be given for boundary conditions
appropriate for $D_{\ell}(z)$.  If apparent-size (angular) distances are desired the reader has
only to compute $D_<(z)=D_{\ell}(z)/(1+z)^2$.  The new solution appears in (\ref{Dell0}),
(\ref{Dellinfty}), (\ref{Dell01}), and
(\ref{Dellinfty1}) expressed in terms of Heun functions $Hl$.  All previously known special
solutions are limiting cases of the general solution (\ref{Dellinfty}) [see
(\ref{hypergeometric}) and Appendix B].  To solve equation (\ref{Af}) it is first rewritten as:
\bea &&(1+z)^3\sqrt{1+\OM z+\OL[(1+z)^{-2}-1]}\times\nonumber\\ &&\hskip 1 in {d\ \over
dz}(1+z)^3\sqrt{1+\OM z+ \OL[(1+z)^{-2}-1]}\,{d\ \over dz}\sqrt{A(z)}\nonumber\\ &&\hskip 2.0 in
+ {(3+\nu)(2-\nu)\over 4}\OM(1+z)^5\sqrt{A(z)}=0.  \label{Ab0} \eea 
This equation is often attributed to Dyer-Roeder (\cite{DC1}, \cite{DC2}) in the literature.
(see Appendix B for some history 
of this equation). 
To date only numerical solutions have been obtained when $\OL\ne 0$, \eg see \cite{AH}, 
\cite{SY}, \cite{KRa}.
It can be  put
into a recognizable form by changing the independent variable from $z$ to $y$ and the dependent
variable from $\sqrt{A(z)}$ to $h$,
\bea 
y&=&y_0(1+z)={\OM\over 1-\OM-\OL}(1+z),\nonumber\\ 
h&=&(1+z)\sqrt{{A\over\delta\Omega}}.
\label{newvariables} 
\eea
The resulting equation is 
\be
{d^2h\over dy^2} +{\left(1+{3\over 2}y\right)y\over y^3+y^2-\bO}\
{dh\over dy} - { {1\over 4}\nu(\nu+1)y+1\over y^3+y^2-\bO}\ h = 0\ , \label{h1} 
\ee
where
$\bO\equiv -\OM^2\OL/(1-\OM-\OL)^3$.  When the cubic $y^3+y^2-\bO=(y-y_1)(y-y_2)(y-y_3)$ is
factored, (\ref{h1}) simplifies to a recognizable form of the Heun equation (see \cite{RA},
\cite{EA}, \cite{WE}, \cite{HK}):  
\be
{d^2h\over dy^2} +\left({\gamma\over y-y_1}+{\delta\over
y-y_2}+{\epsilon\over y-y_3}\right) {dh\over dy} + { \alpha\beta\ y-q\over
(y-y_1)(y-y_2)(y-y_3)}\ h = 0\ , \label{h2} 
\ee
where (\ref{h1}) requires
 \bea
&&\gamma=\delta=\epsilon={1\over 2},\nn &&\alpha = -{1\over 2}\nu,\nn &&\beta= {1\over
2}(\nu+1),\nn &&q=1,
 \label{exponents}  
\eea 
and additionally the three roots to be constrained by:  \bea
&&y_1y_2y_3=\bO=-\OM^2\OL/(1-\OM-\OL)^3,\nonumber\\ &&y_1+y_2+y_3=-1,\nonumber\\
&&y_1y_2+y_1y_3+y_2y_3=0.  \label{rootconstraints} \eea

The Heun equation is slightly more complicated than the hypergeometric equation; it possesses
four regular singular points in the entire complex plane rather than just three.  In the form
given by (\ref{h2}) one of the two exponents of each finite singular point $(y_1,y_2,y_3)$
vanishes and the other exponent is given respectively by
$(1-\gamma,1-\delta,1-\epsilon)$.\footnote[4]{ Recall that an exponent gives the analytic
behavior of a solution within the neighborhood of a regular singular point, \eg
$h=(y-y_1)^{1-\gamma}(1+ c_1(y-y_1)+ \cdots)$.}  The point at $\infty$ is the fourth singular
point and its exponents are $\alpha$ and $\beta$.  For the point at $\infty$ to also be regular
(\ie for this to be a Heun equation) all exponents must sum to a value of 2, equivalently \be
\alpha+\beta+1=\gamma+\delta+\epsilon.  \label{exponentsum} 
\ee
For (\ref{h1}) this necessary
constraint is satisfied.  From (\ref{rootconstraints}) it follows that at least one root has to
be real and complex roots must come in conjugate pairs.  For convenience $y_1$ will be
chosen as real throughout.  This Heun equation (\ref{h2}) is conveniently expressed in terms of a
Riemann P-symbol as:  
\be
P\left\{ \begin{array}{cccccc} y_1 & y_2 & y_3 & \infty \\ 0 & 0 & 0 &
\alpha & y & q \\ 1-\gamma & 1-\delta & 1-\epsilon & \beta\\ \end{array} \label{Py} \right\}.
\ee
The first 4 columns of (\ref{Py}) are the 4 regular singular points and their 2 exponents,
and the 5th column is the independent variable, all analogous to the Riemann P-symbol for the
hypergeometric equation.  The 6th column is the constant $q$ from the numerator of the
coefficient of $h$ in the Heun equation [when put into the standard form of  (\ref{h2})].
The hypergeometric equation is uniquely specified by information about its regular singular
points but Heun requires the extra parameter $q$.  Because the three finite singular points 
of this Heun equation have
values of 1/2 for their nonvanishing exponents,  (\ref{h1}) can be transformed
into the Lame$^{\prime}$ equation.  In this paper, solutions of (\ref{h1}) will be given as local
Heun functions and in a following paper they will be expressed as Lame$^{\prime}$ functions.
When $\Lambda=0$ ($\Rightarrow \bO=0$) equation (\ref{h1}) has only three regular singular points
\ie this Heun equation simplifies to an equation of the hypergeometric type.  Additionally the
corresponding Lame$^{\prime}$ equation reduces to the associated Legendre equation.  Solutions
for $\Lambda =0$ can be written either as combinations of hypergeometric functions or as
associated Legendre functions [see (\ref{hypergeometric}) and (\ref{Legendre}) below].  To
motivate the form of the $\OL\ne 0$ solution, the $\OL=0$ solution will be given first [\cite{KVB} and
\cite{SS}],
\bea 
&&D_{\ell}(\OM=\OO,\OL=0, \nu ; z)\nonumber\\ 
&&= {c\over H_0}{1\over
(\nu+{1\over2})}\times \nonumber\\ 
&& \Biggl[\!\Biggl[ (1+\OO z)^{1+\nu/2}\ \ {}_2F_1\left({\nu\over2}+2,
{\nu\over2}+{3\over 2}; \nu+{3\over2} ; 1-\OO\right) {}_2F_1\left(-{\nu\over2}-1,
-{\nu\over2}-{1\over2}; {1\over2}-\nu; {1-\OO\over 1+\OO z}\right) \nonumber\\ 
&-&{(1+z)^2\over (1+\OO z)^{3/2+\nu/2} } \ {}_2F_1\left(-{\nu\over2}-1, -{\nu\over2}-{1\over2}; {1\over2}-\nu;
1-\OO\right) {}_2F_1\left({\nu\over2}+2, {\nu\over2}+{3\over 2}; \nu+{3\over2} ; {1-\OO\over
1+\OO z}\right) \Biggr]\!\Biggr].\nonumber\\ \label{hypergeometric} 
\eea 
The expected form of the $\OL\ne
0$ solution follows the above where the hypergeometric functions ${}_2F_1$ are replaced by local
Heun functions $Hl$ [see (\ref{Dell0}) and (\ref{Dellinfty})].

From (\ref{rootconstraints}) the three singular points $(y_1,y_2,y_3)$ are chosen from the six
permutations of the three roots:  
\bea 
Y_1&=&-{1\over 3}\left[1-{1\over
v_+}-v_+\right],\nonumber\\ 
Y_2&=&-{1\over 3}\left[1+{1\over v_-}+v_-\right],\nonumber\\
Y_3&=&-{1\over 3}\left[1+{e^{-i{\pi\over 3}}\over v_+}+ e^{i{\pi\over 3}}v_+\right],
\label{yroots} 
\eea 
where 
\bea 
v_+&\equiv& \left[-1+b+\sqrt{b(b-2)}\right]^{1\over
3},\nonumber\\ v_-&\equiv& \left[1-b+\sqrt{b(b-2)}\right]^{1\over 3}, 
\label{yroots1}
\eea 
with 
\be
b\equiv {27\over 2}\bO=-{27\over 2}\OM^2\OL/(1-\OM-\OL)^3.
\label{b}
\ee

The
locations of the three finite singular points $(y_1,y_2,y_3)$ are determined by the value of the
single parameter $b,\ (-\infty<b<\infty)$.  Some important values are shown as contours
in Figure 2.

The standard form for the Heun equation ordinarily has its singularities at $(0,1,a,\infty)$.
The simple linear transformation 
\be
\zeta={y-y_1\over y_2-y_1}, \label{zeta} 
\ee
moves \bea y_1
&\rightarrow &0,\nonumber\\ y_2 &\rightarrow& 1,\nonumber\\ y_3 &\rightarrow& a={y_3-y_1\over
y_2-y_1}={y_1(2+3y_1)\over (y_2-y_1)^2} ={(y_3-y_1)^2\over y_1(2+3y_1)}.  \label{zetaroots} \eea
The latter two equalities are consequences of the useful identity:  \be
(y_2-y_1)(y_3-y_1)=y_1(2+3y_1), \label{rootidentity} 
\ee
which results from
(\ref{rootconstraints}).  In terms of the new variable $\zeta$, (\ref{h1}) becomes 
\be
{d^2h\over
d\zeta^2} +{1\over2}\left({1\over \zeta}+{1\over \zeta-1}+{1\over \zeta-a}\right) {dh\over
d\zeta} + { (-{1\over 2}\nu){1\over 2}(\nu+1)\zeta-q\over \zeta(\zeta-1)(\zeta-a)}\ h = 0\ ,
\label{h3} 
\ee
and the value of $q$ changes to:  
\be
q={1+{1\over4}\nu(\nu+1)y_1\over y_2-y_1}.
\label{q} 
\ee
The new Riemann P-symbol is:

\be
P\left\{ \begin{array}{cccccc} 0 & 1 & a & \infty \\ 0 & 0 & 0 & -{\nu\over2} & \zeta &
 {1+{1\over4}\nu(\nu+1)y_1\over y_2-y_1} \\ {1\over 2} & {1\over 2} & {1\over 2} & {\nu+1\over
 2}\\ \end{array} \label{Pzeta} \right\}.  
\ee
See the figures in Figure 3 for locations of $a$ in
 the complex plane and the trajectories of $\zeta(z)$ starting with $\zeta_0$ (the value of
 $\zeta$ at zero redshift [see (\ref{newvariables})  and (\ref{zeta})]) for the following three cases:  
\bea
 &&b < 0 \longrightarrow y_1= {\rm real,\ } y_2= \bar{y}_3, {\rm \ and\ } |a|=1,\nonumber\\
 &&0\le b \le 2 \longrightarrow y_1,y_2,y_3, a {\rm\ are\ all\ real},\nonumber\\ 
&&2< b
 \longrightarrow y_1= {\rm real,\ } y_2= \bar{y}_3, {\rm \ and\ } |a|=1.  \label{beta} \eea Figure 3 
 gives the proper choices for the three roots $(y_1,y_2,y_3)$ from the six possible orderings
 of $(Y_1,Y_2,Y_3)$ in each of the three $b$ domains.  It also contains values for $a, q$,
 the new variables $\zeta$, and $\zeta_0$.  Hyperbolic and trigonometric variables, $\xi$ and
 $\phi$, can be used to parameterize the values of the three roots (rather than $b$) and they
 are also given in Figure 3.

Boundary conditions on $h$ come directly from its definition (\ref{newvariables}) and the desired
boundary conditions on $\sqrt{A}$ [see (\ref{Aboundary})], 
\bea \sqrt{A}|_s&=&0 \Longrightarrow
h(\zeta_s)=0,\nonumber\\ 
{d\sqrt{A}\over dz}\Bigg|_s &=& -\sqrt{\delta\Omega} {c\over H_s(1+z_s)}
\Longrightarrow {d h\over d z}\Bigg|_s =-{c\over H_s}.  \eea 
Equation (\ref{Dl}) then relates
$D_{\ell}$ to the value of $h$ at the observer, 
\be
D_{\ell}(z_s)=(1+z_s)h(z=0).  \label{Dlh0}
\ee
Using these boundary conditions on two independent solutions $h_1\ \&\ h_2$ of (\ref{h3})
gives 
\be
 h(\zeta)=-{h_1(\zeta_s) h_2(\zeta)-h_2(\zeta_s) h_1(\zeta) \over h_1(\zeta_s) \dot
h_2(\zeta_s)-h_2(\zeta_s) \dot h_1(\zeta_s)} \left[ \left({c\over H_s}\right) \biggr/ {d\zeta\over
dz}\Bigg|_{z_s} \right], \label{h12} 
\ee
 where $\dot h \equiv {d h\over d\zeta}$.  From
(\ref{zeta}) and (\ref{newvariables}) 
\bea 
&&\zeta={y_0(1+z)-y_1\over y_2-y_1},\nn
&&\zeta_0={y_0-y_1\over y_2-y_1},\nn 
&&{d\zeta\over dz}={y_0\over y_2-y_1}.  
\label{zeta0} 
\eea
The denominator of $h(\zeta)$ in (\ref{h12}) can be evaluated using the Wronskian of (\ref{h3}),

\be
h_1(\zeta) \dot h_2(\zeta)-h_2(\zeta) \dot h_1(\zeta)= (C_W){1\over
\sqrt{\zeta(\zeta-1)(\zeta-a)}}, \label{Wronskian} 
\ee
where $C_W$ is a constant.  The square
root in this term can be evaluated using \bea
\sqrt{\zeta(\zeta-1)(\zeta-a)}&=&{\sqrt{y^3+y^2-\bO}\over (y_2-y_1)^{3/2}},\nn
\sqrt{y^3+y^2-\bO}&=&(1+z){y_0^{3/2}\over \sqrt{\Omega_m}} \sqrt{1+\OM z+\OL[(1+z)^{-2}-1]}.
\label{} 
\eea 
With $D_{\ell}$ from (\ref{Dlh0}) and $H_s$ from (\ref{Hs}), equations (\ref{h12})
and (\ref{Wronskian}) give the desired result:  
\be
D_{\ell}(z_s)=-{c(1+z_s)y_0^{1/2}\over
H_0\sqrt{\Omega_m}(y_2-y_1)^{1/2}(C_W)} \left[h_1(\zeta_s) h_2(\zeta_0)-h_2(\zeta_s)
h_1(\zeta_0)\right].  \label{Dell} 
\ee
Figure 2 shows domains in the $(\OM,\OL)$ plane separated
by $b =2$, $b =\infty$, and $ |\zeta_0|=1$.  The $b =\infty$ contour is equivalent to
$\OO=1$.  These contours are important because they separate domains for which different choices
of the two independent solutions $h_1(\zeta)$ and $h_2(\zeta)$ must be taken.  The $ |\zeta_0|=1$
contour divides the $ |\zeta_0|<1$ domain where solutions about the singular point $\zeta=0$ are
chosen from the $ |\zeta_0|>1$ domain where solutions about $\infty$ are chosen.  These choices
are necessary for convergence of the local Heun functions.  For the $\OL=0$ case, only 
analytic expressions 
about $\infty$ were required [see (\ref{hypergeometric})].

The Heun Function $Hl(a,q;\alpha,\beta,\gamma,\delta;\zeta)$ is the analytic solution of
(\ref{h2fn}) defined by the infinite series (\ref{Hseries}), see \cite{RA}.  It converges in a
circle centered on $\zeta=0$ which extends to the nearest singular point $1$ or $a$.  This
solution is analogous to the ${}_2F_1$ solution of the hypergeometric equation but unfortunately
does not appear in any of the common computer libraries.  When $c_0=1$ is chosen (as will be done
here) the series is:  
\be
Hl(a,q;\alpha,\beta,\gamma,\delta;\zeta) \equiv 1+ \sum_{r=1}^{\infty}
c_r\zeta^r, \label{Hseries} 
\ee
where the $c_r$ are constrained by a three term recursion
relation (take $c_{-1}=0$):  
\be
P_rc_{r-1}-(Q_r+q)c_r+R_rc_{r+1}=0, \label{recursion} 
\ee
with
\bea &&P_r\equiv (r-1+\alpha)(r-1+\beta),\nn 
&&Q_r\equiv
r[(r-1+\gamma)(1+a)+a\delta+\epsilon],\nn 
&&R_r\equiv(r+1)(r+\gamma)a.  \label{recursionCons} 
\eea
The $\epsilon$ parameter is not included as an argument in
$Hl(a,q;\alpha,\beta,\gamma,\delta;\zeta)$ because of the constraint (\ref{exponentsum}).  This
series corresponds to the zero exponent for the regular singular point $\zeta=0$ and will be
taken as $h_1(\zeta)$ in (\ref{Dell}) when $|\zeta_0|<1$.  The second independent solution is \be
h_2(\zeta)= \zeta^{1-\gamma}Hl(a,q_{II};\alpha_{II},\beta_{II},\gamma_{II},\delta;\zeta),
\label{h2fn} 
\ee
where four parameters have changed, \bea &&q_{II}\equiv
(a\delta+\epsilon)(1-\gamma)+q,\nn &&\alpha_{II}\equiv \alpha+1-\gamma,\nn &&\beta_{II}\equiv
\beta+1-\gamma,\nn &&\gamma_{II}\equiv 2-\gamma.  \eea

The constant $C_W$ in the Wronskian can be evaluated for the $\zeta\sim 0$ expansion using $h_1$
and $h_2$ above, 
\be
C_W={1\over 2}\sqrt{a}.  
\ee
This gives an expression for the luminosity
distance appropriate for $|\zeta_0|< 1$, 
\bea 
&&D_{\ell}(\OM,\OL,\nu;z)= -{c(1+z)\over H_0{1\over
2}\sqrt{\Omega_m}}\sqrt{{y_0(y_0-y_1)\over y_1(2+3y_1)}}\times\nn 
&&\Biggl[\!\Biggl[
Hl\left(a,{1+{1\over4}\nu(\nu+1)y_1\over\sqrt{y_1(2+3y_1)}}\sqrt{a}; -{\nu\over 2},{\nu+1\over
2},{1\over 2},{1\over 2};{y_0(1+z)-y_1\over \sqrt{y_1(2+3y_1)}}\sqrt{a}\right)\nn 
&&\times Hl\left(a,{3+(\nu^2+\nu-3)y_1\over 4 \sqrt{y_1(2+3y_1)}}\sqrt{a}; -{\nu-1\over
2},{\nu+2\over 2},{3\over 2},{1\over 2};{y_0-y_1\over \sqrt{y_1(2+3y_1)}}\sqrt{a}\right) \nn
&&-\sqrt{{y_0(1+z)-y_1 \over y_0-y_1}} Hl\left(a,{3+(\nu^2+\nu-3)y_1\over 4
\sqrt{y_1(2+3y_1)}}\sqrt{a}; -{\nu-1\over 2},{\nu+2\over 2},{3\over 2},{1\over
2};{y_0(1+z)-y_1\over \sqrt{y_1(2+3y_1)}}\sqrt{a}\right)\nn 
&&\times
Hl\left(a,{1+{1\over4}\nu(\nu+1)y_1\over \sqrt{y_1(2+3y_1)}}\sqrt{a}; -{\nu\over 2},{\nu+1\over
2},{1\over 2},{1\over 2};{y_0-y_1\over \sqrt{y_1(2+3y_1)}}\sqrt{a}\right)
 \Biggr]\!\Biggr], \label{Dell0}
\eea 
where the source redshift $z_s$ has again been replaced by $z$.  The required values of $y_1$ and $a$ can be
found in Figure 3 for all three $b$ domains and $y_0=\OM/(1-\OO)$ is the value of $y$ at $z=0$
[see (\ref{newvariables})].  Even though the above Heun functions contain complex arguments and
parameters, they are real valued functions of the real redshift varable $z$.  As soon as these
functions become available in Mathematica, expressions (\ref{Dell0}) and (\ref{Dellinfty}) will
be immediately useful.  Untill then, simpler expansions suitable for $|a|=1$ are indicated in
Appendix C. 

The solution similar to (\ref{Dell0}) but suitable for $|\zeta_0|>1$ is given by choosing \be
h_1(\zeta)=\zeta^{-\alpha}Hl\left({1\over a},\underline{q};
\alpha,\underline{\beta},\underline{\gamma},\delta;{1\over \zeta}\right), \label{h1b} 
\ee
and 
\be
h_2(\zeta)=\zeta^{-\beta}Hl\left({1\over a},\underline{q}_{II};
\underline{\alpha}_{II},\underline{\beta}_{II},\underline{\gamma}_{II},\delta;{1\over \zeta}\right),
\label{h2b} 
\ee
where seven parameters have now changed:  
\bea 
&&\underline{q}\equiv{q\over
a}-\alpha\left[\beta\left(1+{1\over a}\right) -{\delta\over a}-\epsilon\right]\nn
&&\underline{\beta}\equiv-\beta+\delta+\epsilon\nn &&\underline{\gamma}\equiv1+\alpha-\beta\nn
&&\underline{q}_{II}\equiv \underline{q}+\left({\delta\over a}+\epsilon\right)
\left(1-\underline{\gamma}\right),\nn &&\underline{\alpha}_{II}\equiv
\alpha+1-\underline{\gamma},\nn &&\underline{\beta}_{II}\equiv
\underline{\beta}+1-\underline{\gamma},\nn &&\underline{\gamma}_{II}\equiv 2 -\underline{\gamma}.
\label{bar} 
\eea
 For this choice of $h_1$ and $h_2$ the constant in the Wronskian
(\ref{Wronskian}) becomes 
\be
C_W=-\beta+\alpha=-(\nu+{1\over 2}).  
\ee
From (\ref{Dell}) the
resulting expression for the luminosity distance is

\bea 
&&D_{\ell}(\OM,\OL,\nu;z)={c(1+z)\over
H_0(\nu+{1\over 2})\sqrt{\Omega_m}\sqrt{(1+z)-y_1/y_0}}\times\nn 
&&\Biggl[\!\Biggl[
 \left({y_0(1+z)-y_1
\over y_0-y_1}\right)^{{\nu+1\over 2}}\nn 
&&\hskip .25 in \times Hl\left({1\over
a},{4-\nu^2+\nu(1-2\nu)y_1\over 4\sqrt{y_1(2+3y_1)}}{1\over\sqrt{a}}; -{\nu\over 2},{1-\nu\over
2},{1-2\nu\over 2},{1\over 2}; {\sqrt{y_1(2+3y_1)}\over y_0(1+z)-y_1}{1\over\sqrt{a}}\right)\nn
&&\hskip .25 in \times Hl\left({1\over a}, {(3+\nu)(1-\nu)-(3+2\nu)(1+\nu)y_1\over 4
\sqrt{y_1(2+3y_1)}}{1\over\sqrt{a}}; {1+\nu\over 2},{\nu+2\over 2},{3+2\nu\over 2},{1\over 2};
{\sqrt{y_1(2+3y_1)}\over y_0-y_1}{1\over\sqrt{a}}\right) \nn 
&&-\left({y_0-y_1 \over
y_0(1+z)-y_1}\right)^{{\nu\over 2}}\nn 
&&\hskip .25 in \times Hl\left({1\over a},
{(3+\nu)(1-\nu)-(3+2\nu)(1+\nu)y_1\over 4 \sqrt{y_1(2+3y_1)}}{1\over\sqrt{a}}; {1+\nu\over
2},{\nu+2\over 2},{3+2\nu\over 2},{1\over 2}; {\sqrt{y_1(2+3y_1)}\over
y_0(1+z)-y_1}{1\over\sqrt{a}}\right)\nn 
&&\hskip .25 in \times Hl\left({1\over
a},{4-\nu^2+\nu(1-2\nu)y_1\over 4\sqrt{y_1(2+3y_1)}}{1\over\sqrt{a}}; -{\nu\over 2},{1-\nu\over
2},{1-2\nu\over 2},{1\over 2}; {\sqrt{y_1(2+3y_1)}\over y_0-y_1}{1\over\sqrt{a}}\right) \Biggr]\!\Biggr].
\label{Dellinfty} 
\eea
The  special case of $\OO\equiv\OM+\OL = 1$ can be obtained from (\ref{Dell0}) and (\ref{Dellinfty})
 by taking the appropriate limits. Some details of this process along with the resulting
luminosity distance are given in Appendix A. 
For those values of $\OM$ and $\OL$ where $|\zeta_0|< 1$
it is clear that for large enough values of $z$, $|\zeta|> 1$ and  hence (\ref{Dell0}) is no longer
valid ($Hl$ no longer converges).  For some values in the $(\OM,\OL)$ plane above the
$|\zeta_0|= 1$ contour in Figure  2, (\ref{Dell0}) will not converge for a SNe Ia range of $z\sim 0.5$,
but for most values it does.

In the next section several plots of magnitude vs.  redshift are made to illustrate the importantce
of take clumping into account when attempting to
determine $\OM$ and $\OL$.
\section{\mz PLOTS FOR CLUMPY UNIVERSES \& CONCLUSIONS} \label{sec-mzplots} In this section several
magnitude-redshift plots are given to illustrate the effects that density clumps can have on the
\mz relation and consequently on a determination of $\OM$ and $\Lambda$ made by using this
relation.  Because \mz depends differently on $\OM$ and $\OL$ as a function of redshift for the
FLRW models, both parameters could in principle be determined from a sufficient quantity of
accurate SNe Ia data.  Clumping provides an additional parameter $\nu$ which complicates 
any such determination.
As can be seen
from (\ref{Dlseries}) the dependence of \mz on this additional parameter could also be 
determined
by enough data.  However, such a triple determination is certainly more complicated.  
What will be
done here to illustrate the effects of the $\nu$ parameter is to plot multiple \mz curves for
various values of all three parameters $\nu, \ \OM$, and $\OL$.  
In all plots the unit of distance is taken to be $c/H_0$.
In these figures 
$D_{\ell}$ is plotted on a magnitude scale, 5 Log $D_{\ell}$ (\ie the distance modulus plus 
5\,Log $10pc\, H_0/c$).
In Figure 4, $\OL$ is held
fixed while $\nu$ and $\OM$ are varied and in Figure 5, $\OM$ is held fixed while $\nu$ and
$\OL$ are varied.  In Figure 6, $\OO=\OM+\OL=1$ is fixed while all three parameters
vary.

In Figure 7 the sensitivity of observed magnitudes to variations of $\OM$ is illustrated 
by fixing $z=0.83$ and $\OL=0.1$.
In Figure 8 a similar plot is given showing the sensitivity  to variations of $\OL$. 
The importance of the clumping parameter is easily seen from these last two figures. 
If the distance modulus of a source such as SN 1997ap at $z=0.83$  were precisely known 
(\eg see the two sample horizontal lines in Figure 7) then  a determination of  $\OM$ 
could be made, 
assuming $\OL$ were somehow known. Likewise, from Figure 8, a determination of $\OL$ could be made 
if $\OM$ were somehow known.  From Figure 7 the reader can easily see that
the determined value of $\OM$ depends on the clumping parameter $\nu$. The $\OM$ value 
will be about 95\% larger for a $\nu=2$ completely clumpy universe than it will be 
for a $\nu=0$ completely smooth FLRW  universe. Equivalently, $\OM$ could be 
underestimated by as much as 50\% if the FLRW is used. The maximum underestimate is reduced to  33\% 
 at the smaller redshift of $z=0.5$ (see a similar result for $\OL=0$ in \cite{KVB}).
These conclusions are not sensitive to the value of $\OL$.

Slightly different conclusions follow from  Figure 8 about $\OL$. 
The discrepancy in the determined value of $\OL$ is $\Delta \OL \sim -0.14$   
for $\nu=2$ compared to $\nu=0$, and is not sensitive to the 
distance modulus.  The  discrepancy is halved, $\Delta \OL \sim -0.07$, at 
a smaller redshift of $z=0.5$.

A minimal estimate  of the quantity of data  required to begin  distinguishing 
between the various $\nu$ values can easily be made. At 
$z=0.5$  the differences in observed magnitudes of a SN Ia in a $\nu=0$ (100\% smooth FLRW) 
and a $\nu=2$ (100\% clumpy) 
universe is about $\Delta m \sim 0.02$ if $\OM\sim 0.2$, and $\Delta m \sim 0.09$ if $\OM\sim 0.8$. 
These differences are not sensitive to  $\Lambda$.
With corrected-intrinsic and observed magnitude uncertainties of $\pm 0.2$,\ \cite{BD}, 
data on over 200 SNe Ia will be required if we live in a low density universe and over a dozen 
if we live in a higher density one.  

The results presented here (\ref{Dell0}),(\ref{Dellinfty}),(\ref{Dell01}), and (\ref{Dellinfty1}) 
for the `intergalactic' distance--redshift relation are quite general. They contain  
corrections (for mass 
inhomogeneities) to the standard FLRW result, applicable to observations where
gravitational lensing can be neglected, \ie observations where the conformal (Weyl)
curvature doesn't produce significant average shear in (\ref{Af}). Even though the original 
area equation (\ref{Af}) was rigorously established for a particular type of Swiss cheese model, 
the resulting equation which neglects lensing (\ref{Ab0}) is expected to be widely applicable
to observations at redshifts of $z=1$ and less. 
Application of its solution to a given set of observations requires that 
the average fraction of the mass density contained in the observing 
beams (\ie the $\nu$ parameter)  be determined. This fraction obviously depends on the 
number as well as the type of object observed.    Collecting CMB radiation at wide angles is 
likely to produce a $\nu=0$ value but observing a few dozen SNe Ia might well result 
in a value close to $\nu=2$ (\ie we might in fact live in a universe where mass, dark or otherwise, is primarily 
associated with galaxies).  If a significant fraction of 
the universe's mass density is 
clumped on galactic scales, then the effects of these clumps on SNe Ia observations should be
taken into account by using the lensing formulas rather than by  decreasing $\nu$
to zero. 
Recent numerical work by \cite{HD} confirms the assertion that, given galaxy clumping, 
the cross-sectional area 
of a \underbar{typical} light beam will not follow the FLRW area-redshift relation.  
Instead the area will follow more closely one of the `intergalactic' \mz relations given here, until 
a lensing event occurs.  
The new luminosity distances presented  here
represent the theoretical minimum of the observed magnitudes and are especially
applicable to situations where lensing is infrequent (\ie where the most probable value is closer 
to the min than the mean). Because \cite{HD} did not include any diffuse 
transparent matter, the  applicable  \mz  relations given here are those with $\nu=2$.
For $\Lambda=0$ and $\Omega_0<1$ it is  the Dyer-Roeder solution (\ref{DC1}) 
and for $\Lambda=0$ and $\Omega_0=1$
its the $\nu=2$ solution of Dashevskii \& Slysh (\ref{Dash}).

The $\nu=0$ (standard FLRW) result represents the theoretical `mean' for  \mz for a universe 
in which only weak-lensing events occur. For extremely non-symmetric probability distributions, 
the ``mean" 
is not likely the best estimator - in this case the ``most probable" is likely  better, \cite{SD}.
\acknowledgements
The author would like to thank Tamkang University for their kind hospitality and support
during an extended visit to Taiwan in the Spring of '97 where this work was first presented. 
The author would also like to thank D. Branch and E. Baron for suggesting changes in the final draft.
\appendix
\section{THE SPECIAL CASE: $\OO=\OM+\OL=1$} \label{sec-Appendix A} 
A complete derivation of the $\OO=\OM+\OL=1$ case can be done by  
introducing a new independent variable in (\ref{h1}), $y\rightarrow -y/\Delta$ 
where $ \Delta\equiv \OM+\OL -1$ and then taking the  limit $\Delta \rightarrow 0$. 
The resulting differential equation which replaces (\ref{h1}) has the same exponents given 
in (\ref{exponents}) but has $q=0$. The  three 
finite regular singular points are now located at
$(y_1,y_2,y_3)=(\OM^2\OL)^{1/3}(-1,e^{-i\pi/3},e^{i\pi/3})$. When the modified
 equation (\ref{h1}) is transformed by (\ref{zeta}) 
a modified (\ref{h3}) results which is described by  the Riemann P-symbol:
\be
P\left\{ \begin{array}{cccccc} 0 & 1 & e^{i\pi/3} & \infty \\ 0 & 0 & 0 & -{\nu\over2} & 
\left[\left({\OM\over\OL}\right)^{1\over 3}(1+z)+1\right]{e^{i\pi/6}\over\sqrt{3}}, &
 -{\nu(\nu+1)\over4\sqrt{3}}e^{i\pi/6} \\ {1\over 2} & {1\over 2} & {1\over 2} & {\nu+1\over
 2}\\ \end{array} \label{Pzeta1} \right\}.  
\ee
The solutions to this simpler Heun equation with boundary conditions appropriate for  
luminosity distance $D_{\ell}$ can be   obtained directly from  
(\ref{Dell0}) and (\ref{Dellinfty}) by
simply substituting $y_0=\OM/\Delta,\ y_1=-(\OM^2\OL)^{1/3}/\Delta, a=e^{i\pi/3}$,  
and then taking the limit $\Delta \rightarrow 0$. 
This gives an expression for the luminosity
distance, appropriate for $|\zeta_0|< 1$, 
\bea
 &&D_{\ell}(\OM,\OL,\nu;z)=
-{
c(1+z)\sqrt{1+({\OL\over\OM})^{1\over3}}
\over
 H_0{1\over2}\sqrt{3}\OM^{1\over6}\OL^{1\over3}
}\times\nn 
&&
\Biggl[\!\Biggl[
Hl\left(
e^{i{\pi\over 3}},
-{\nu(\nu+1)\over4\sqrt{3}}e^{i\pi/6};
 -{\nu\over 2},{\nu+1\over2},{1\over 2},{1\over 2};
\left[\left({\OM\over\OL}\right)^{1\over 3}(1+z)+1\right]{e^{i\pi/6}\over\sqrt{3}}
\right)
\nn &&\times 
Hl\left(
e^{i{\pi\over 3}},
-{(\nu^2+\nu-3)\over4\sqrt{3}}e^{i\pi/6};
 -{\nu-1\over2},{\nu+2\over 2},{3\over 2},{1\over 2};
\left[\left({\OM\over\OL}\right)^{1\over 3}+1\right]{e^{i\pi/6}\over\sqrt{3}}
\right) \nn
&&-\sqrt{{(1+z)+({\OL\over\OM})^{1\over3} \over 1+({\OL\over\OM})^{1\over3}}} 
\nn &&\times 
Hl\left(
e^{i{\pi\over 3}},
-{(\nu^2+\nu-3)\over4\sqrt{3}}e^{i\pi/6};
 -{\nu-1\over 2},{\nu+2\over 2},{3\over 2},{1\over2};
\left[\left({\OM\over\OL}\right)^{1\over 3}(1+z)+1\right]{e^{i\pi/6}\over\sqrt{3}}
\right)
\nn &&\hskip .5 in \times
Hl\left(
e^{i{\pi\over 3}},
-{\nu(\nu+1)\over4\sqrt{3}}e^{i\pi/6};
 -{\nu\over 2},{\nu+1\over2},{1\over 2},{1\over 2};
\left[\left({\OM\over\OL}\right)^{1\over 3}+1\right]{e^{i\pi/6}\over\sqrt{3}}
\right) \Biggr]\!\Biggr], \label{Dell01}
\eea
The expression for the luminosity
distance appropriate for $|\zeta_0|> 1$ is obtained by applying the limiting proceedure to 
(\ref{Dellinfty}),
\bea 
&&D_{\ell}(\OM,\OL,\nu;z)=
{
c(1+z)\over
H_0(\nu+{1\over 2})\sqrt{\Omega_m}\sqrt{(1+z)+({\OL\over\OM})^{1\over3}}
}
\times\nn 
&&\Biggl[\!\Biggl[ 
\left(
{(1+z)+({\OL\over\OM})^{1\over3}\over 
1+({\OL\over\OM})^{1\over3}}
\right)^{{\nu+1\over 2}}\nn 
&&\hskip .25 in \times 
Hl\left(
e^{-i{\pi\over 3}},
{\nu(2\nu-1)\over4\sqrt{3}}e^{-i\pi/6};
 -{\nu\over 2},{1-\nu\over2},{1-2\nu\over 2},{1\over 2};
  \sqrt{3}\left[\left({\OM\over\OL}\right)^{1\over 3}(1+z)+1\right]^{-1}e^{-i\pi/6}
\right)\nn
&&\hskip .25 in \times 
Hl\left(
e^{-i{\pi\over 3}}, 
{(\nu+1)(2\nu+3)\over4\sqrt{3}}e^{-i\pi/6}; 
{1+\nu\over 2},{\nu+2\over 2},{3+2\nu\over 2},{1\over 2};
\sqrt{3}\left[\left({\OM\over\OL}\right)^{1\over 3}+1\right]^{-1}e^{-i\pi/6}
\right) \nn 
&&-\left({1+({\OL\over\OM})^{1\over3}\over 
(1+z)+({\OL\over\OM})^{1\over3}}\right)^{{\nu\over 2}}\nn 
&&\hskip .25 in \times 
Hl\left(e^{-i{\pi\over 3}},
{(\nu+1)(2\nu+3)\over4\sqrt{3}}e^{-i\pi/6}; 
{1+\nu\over2},{\nu+2\over 2},{3+2\nu\over 2},{1\over 2};
\sqrt{3}\left[\left({\OM\over\OL}\right)^{1\over 3}(1+z)+1\right]^{-1}e^{-i\pi/6}
 \right)\nn 
&&\hskip .25 in \times 
Hl\left(e^{-i{\pi\over 3}},
{\nu(2\nu-1)\over4\sqrt{3}}e^{-i\pi/6};
 -{\nu\over 2},{1-\nu\over2},{1-2\nu\over 2},{1\over 2};
\sqrt{3}\left[\left({\OM\over\OL}\right)^{1\over 3}+1\right]^{-1}e^{-i\pi/6}
 \right) \Biggr]\!\Biggr].
\label{Dellinfty1} 
\eea
These expressions were used to produce Figure 6 of \S \ref{sec-mzplots}. 
The $\OL \rightarrow 0$ limit of (\ref{Dellinfty1}) results in the solution (\ref{Dash}) 
below, first given by \cite{DV}.
  
\section{PREVIOUSLY KNOWN SOLUTIONS FOR \hbox{{\it D}$_{\ell}$}(${\lowercase {z}}$)
WHEN $B_0=0$ } \label{sec-Appendix B} 
Until now,
analytic solutions to the average area equation (\ref{Af}), neglecting lensing (\ie putting
$\langle\xi^2\rangle$=0), have been found only for $\Lambda=0$.  The earliest solutions were written down before
the equation was formulated by \cite{KR}.  The standard homogeneous FRW solution was given by
\cite{MW}.  It is the $\rho_I=0$ (\ie $\nu=0$) solution of (\ref{hypergeometric}) and
(\ref{Legendre}), 
\be
D_{\ell}(\OM=\OO,\OL=0,\nu=0;z) ={2c\over H_0\OO^2}\left\{\OO
z+(\OO-2)\left(\sqrt{1+\OO z}-1\right)\right\}.  \label{Mattig} 
\ee
The $\OO=1$ solution was 
given by \cite{DV}, \footnote[5]{\cite{Zel} had given the $\OM=1,\ \Lambda=0,\ \nu=2$ solution 
along with the first derivation of (\ref{Ab0}) restricted to that particular case. 
Zel'dovich seems to be the first to recognize the importance of inhomogeneities 
on FRW distances and to attempt to make the needed modifications. 
\cite{DVZ} later extended the 
equation to $\Omega_m\ne 1$ but only for $\nu=2$. It was \cite{DV} that produced the first
$\nu\ne 2$ or 0, $\Lambda = 0$, $\OM\ne 1$ version of (\ref{Ab0}) along with the $\OM=1$ solution (\ref{Dash}).}
\be
D_{\ell}(\OM=1,\OL=0,\nu;z) = {c\over H_0}{1\over
(\nu+{1\over2})}\left[(1+z)^{{\nu\over2}+1}- (1+z)^{-{\nu\over2}+{1\over 2}}\right].
\label{Dash} 
\ee
The $\nu=2$ (\ie $\rho_D=0$) solution is due to \cite{DC1} :  \bea
&&D_{\ell}(\OM,\OL=0,\nu=2;z) \nonumber\\ &&\hskip .5 in ={c\over H_0}{\OO(1+z)^2\over
4(1-\OO)^{3/2}}\Biggl[{3\OO\over 2(1-\OO)} \ln\left\{ \left({1+\sqrt{1-\OO}\over1-\sqrt{1-\OO}
}\right) \left({\sqrt{1+\OO z}-\sqrt{1-\OO} \over \sqrt{1+\OO z}+\sqrt{1-\OO} }\right)
\right\}\nonumber\\ &&\hskip .5 in +{3\over\sqrt{1-\OO}}\left({\sqrt{1+\OO z}\over 1+z} -1\right)
+{2\sqrt{1-\OO}\over \OO}\left( 1 - {\sqrt{1+\OO z}\over (1+z)^2}\right)\Biggr]\,.  \label{DC1}
\eea This result can be rewritten using the identity 
\be
\sinh^{-1}\sqrt{{1-\OO\over \OO(1+z)}} =
{1\over2} \ln\left({\sqrt{1+\OO z}+\sqrt{1-\OO} \over \sqrt{1+\OO z}-\sqrt{1-\OO} }\right)\,,
\label{sinh} 
\ee
in the form actually given by \cite{DC1}.  When $\OO > 1$ equation (\ref{DC1})
is analytically continued using $\sqrt{1-\OO} \longrightarrow \pm i \sqrt{\OO-1}$, which
simplifies by using, $\sinh^{-1}(i x) = i \sin^{-1}(x)$ to give a form containing only real
variables, \bea &&D_{\ell}(\OM,\OL=0,\nu=2;z)\nonumber\\ &&\hskip .5 in ={c\over
H_0}{\OO(1+z)^2\over (\OO-1)^{3/2}}\Biggl[{3\OO\over \OO-1} \left\{ \sin^{-1}\sqrt{{2 \OO-1\over
2 \OO}} -\sin^{-1}\sqrt{{2 \OO-1\over 2 \OO(1+z)}} \right\}\nonumber\\ &&\hskip .5 in
+{3\over2\sqrt{\OO-1}}\left({\sqrt{1+\OO z}\over 1+z} -1\right) -{\sqrt{\OO-1}\over \OO}\left( 1
- {\sqrt{1+\OO z}\over (1+z)^2}\right)\Biggr]\,.  \label{DR2} 
\eea 
The above solution was the
first analytic solution to the $\Lambda=0$ version of (\ref{Af})  and 
cited \cite{KR} as the origin of the equation.  This solution represents the most extreme effects that clumps perhaps have on
the standard Mattig result (\ref{Mattig}) and is therefore one of the more interesting
and useful solutions.  For this solution \underline{no} matter converges
the light beam.  The following solution (\ref{DC2}) appeared in \cite{DC2} and after the above solution  but did
not cite \cite{KR}.  Instead, the derivation was essentially 
repeated neglecting lensing (\ie putting the shear term to zero) and assuming, 
without justification, that the FRW relation between redshift and affine parameter was valid. 
Many authors have subsequently
dubbed (\ref{Af}) with $B_0=0$ as the Dyer-Roeder equation. As pointed out in the previous footnote 
several other non-rigorous derivations of FRW versions of (\ref{Ab0}) already existed by 1966.
Besides making frequent and successful use of the equation, Dyer-Roeder's contribution 
to its development was 
to give 
two special solutions to it and to extend the equation to include $\Lambda \ne 0$. 
It is the authors opinion that if anyone deserves to have their name attached to a 
curved-space optics equation it's R. K. Sachs because 
equations such as (\ref{Af}) are direct applications of \cite{SR}.
The $\nu=1$ (\ie $\rho_D=2\rho_I)$ solution of (\ref{Af}) that appeared in \cite{DC2} was: 
\be
D_{\ell}(\OM,\OL=0,\nu=1;z) = {4\over
3\OO^2}\left[{3\over 2}\OO-1+{1\over 2}\OO z \sqrt{1+\OO z} - {3\over 2}\OO-1\right].
\label{DC2} 
\ee

The general $\Lambda=0$ solution (\ref{hypergeometric}) was only recently obtained and can be
written using associated Legendre functions (\cite{KVB} and \cite{SS}) as \bea
&&D_{\ell}(\OM,\OL=0,\nu;z) ={c\over H_0}{1\over\sqrt{1-\OO}} { 2(1+z)\over
(\nu+2)(\nu+1)(\nu)(\nu-1) } \times \nonumber\\ &&
\left[Q^2_{\nu}\left({1\over\sqrt{1-\OO}}\right) P^2_{\nu}\left({\sqrt{1+\OO
z}\over\sqrt{1-\OO}}\right)- P^2_{\nu}\left({1\over\sqrt{1-\OO}}\right)
Q^2_{\nu}\left({\sqrt{1+\OO z}\over\sqrt{1-\OO}}\right)\right].\nonumber\\ \label{Legendre} 
\eea
The general $\Lambda\ne 0$ solutions (\ref{Dell0}),(\ref{Dellinfty}),
(\ref{Dell01}),and (\ref{Dellinfty1}) take on a form similar to this 
except the associated Legendre functions $P^2_{\nu}\ \&\ Q^2_{\nu}$ 
are replaced by Lame$^{\prime}$
functions.
\section{A SIMPLIFICATION FOR $Hl$ AND MATHEMATICA CODE FOR $D_{\ell}(\OM,\OL,\nu;z)$ } \label{sec-Appendix C} 
Because the Heun functions are not yet available as standard computer routines,
some programming skills  are required if expressions 
(\ref{Dell0}),(\ref{Dellinfty}),(\ref{Dell01}), and (\ref{Dellinfty1}) are to be made used of. 
For those interested in computing the Heun functions used in these expressions the modified 
form of (\ref{Hseries}) used to produce Fig's. 4-8 is given and for those simply interested 
in plots of the luminosity distance, a few lines of 
Mathematica code are also given.

For most points in the $\OM$-$\OL$ plane of Figure 2, $Hl$ as given by (\ref{Hseries}) contains 
complex coefficients $C_r$ as well as a complex variable $\zeta$. Because the needed $Hl$ 
are in fact 
real functions of a real variable $z$ it is convenient to modify this series to make its
coeffecients real. The modification 
is slightly different for expansions about $\zeta =0$ and $\zeta = \infty$.
For the expansions about $\zeta =0$ a convenient coefficient to iterate is $\hat C_r$ defined by:
\be
C_r\zeta^r={\hat C_r\over r!(\gamma)_r2^{2r}[(y_3-y_1)(y_2-y_1)]}(y-y_1)^r,
\ee
and for the expansion about $\infty$,
\be
C_r\left({1\over \zeta}\right)^r=
{\hat C_r\over r!(\gamma)_r2^{2r}}\left({1\over y-y_1}\right)^r.
\ee
The new $\hat C_r$ are real polynomials and hence easily evaluated and simplified.

For those who want only to plot \mz relations a Mathematica routine which
numerically integrates (\ref{h1}) with boundary conditions (\ref{Aboundary}) appropriate for
obtaining $(H_0/c)\times D_{\ell}(\OM,\OL,\nu;z)$ follows. As input the program requires values for Om,
 Ol, and nu
which stand for $\OM,\OL,$ and $\nu$ respectively.
\vskip 10pt
\vbox{
\parskip = 1pt
\parindent = 35pt
y0= Om/(1$-$Om$-$Ol)

bO= $-$Om\^{}2*Ol/(1$-$Om$-$Ol)\^{}3

P= (1+3/2y)y/(y\^{}3+y\^{}2$-$bO)

Q= $-$(1+nu*(nu+1)*y/4)/(y\^{}3+y\^{}2$-$bO)

distance= NDSolve[\{d''[y]+P*d'[y]+Q*d[y] == 0, 

\hskip 30pt d[y0]==0,  d'[y0]==1/y0\}, d,  \{y,  y0,  2*y0\}]
 
Plot[Evaluate[(1+z)*d[y0*(1+z)] /. distance], \{z, 0, 1\}]
}
This  routine was used to check for errors in 
(\ref{Dell0}),(\ref{Dellinfty}),(\ref{Dell01}), and (\ref{Dellinfty1}).

\clearpage

\figcaption[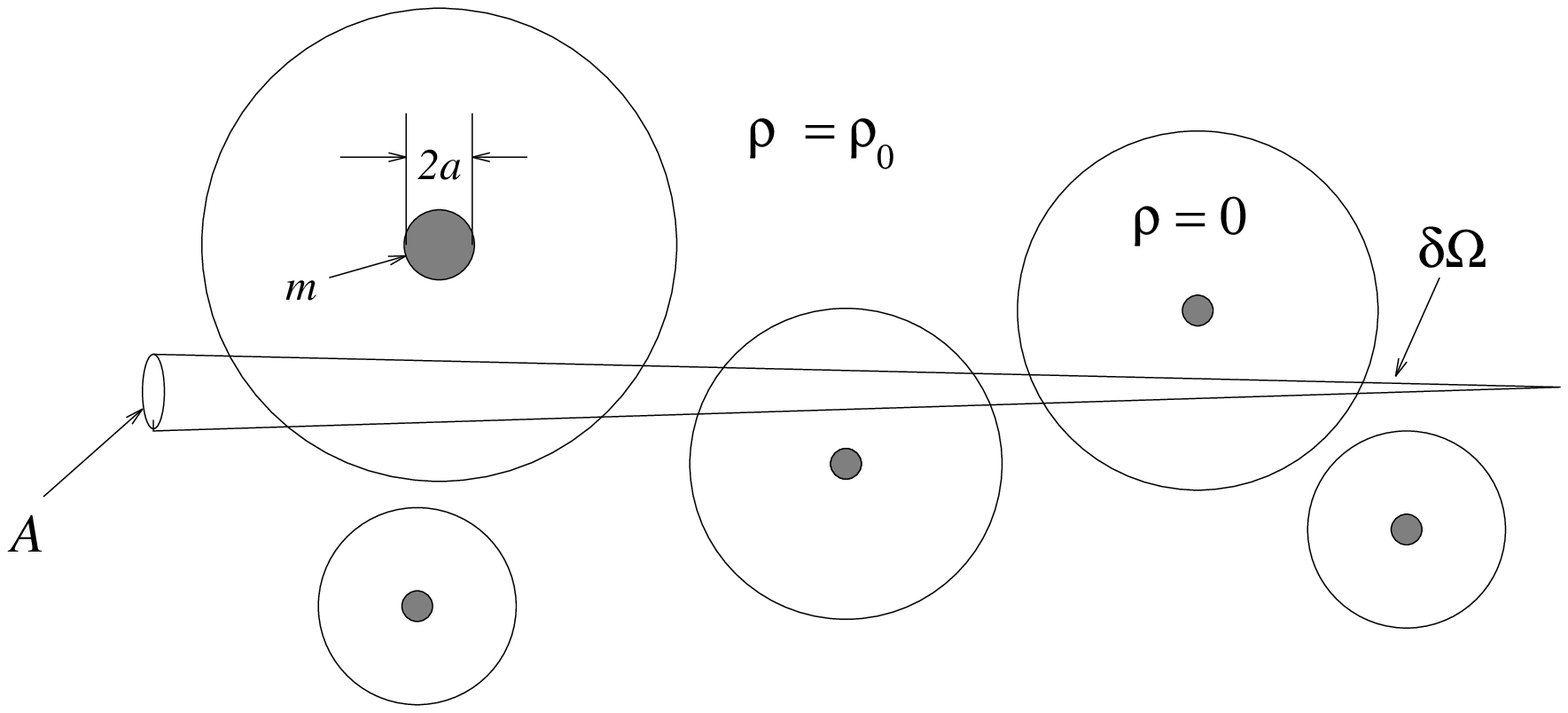]{Radiation beam of cross-sectional area  $A$ propagating 
through a Swiss cheese universe from distant source to observer.
\label{fig1_8}}

\figcaption[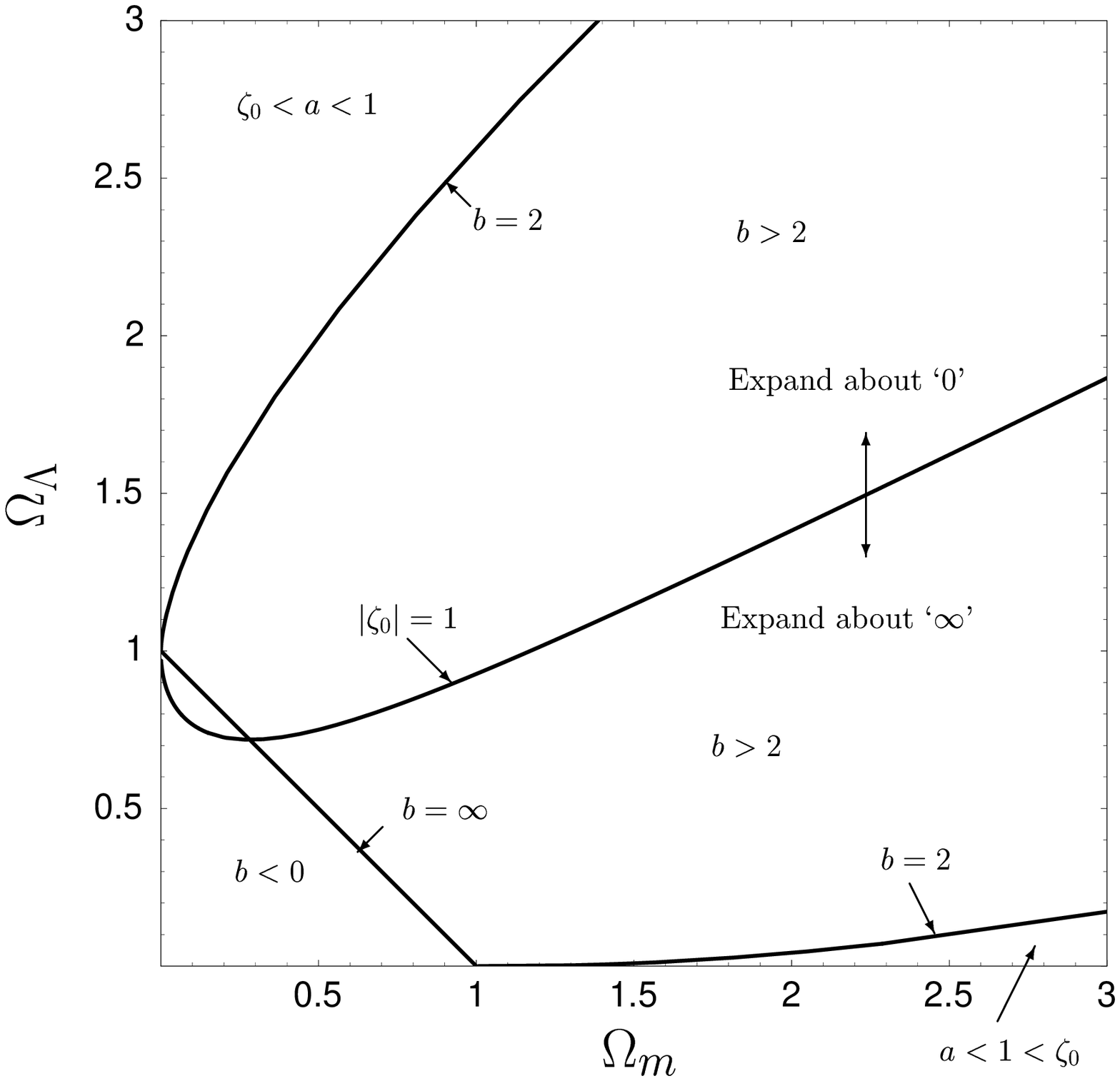]{The $\Omega_m$-$\OL$ plane and various domains required for 
the needed Heun functions.
\label{fig2_8}}

\figcaption[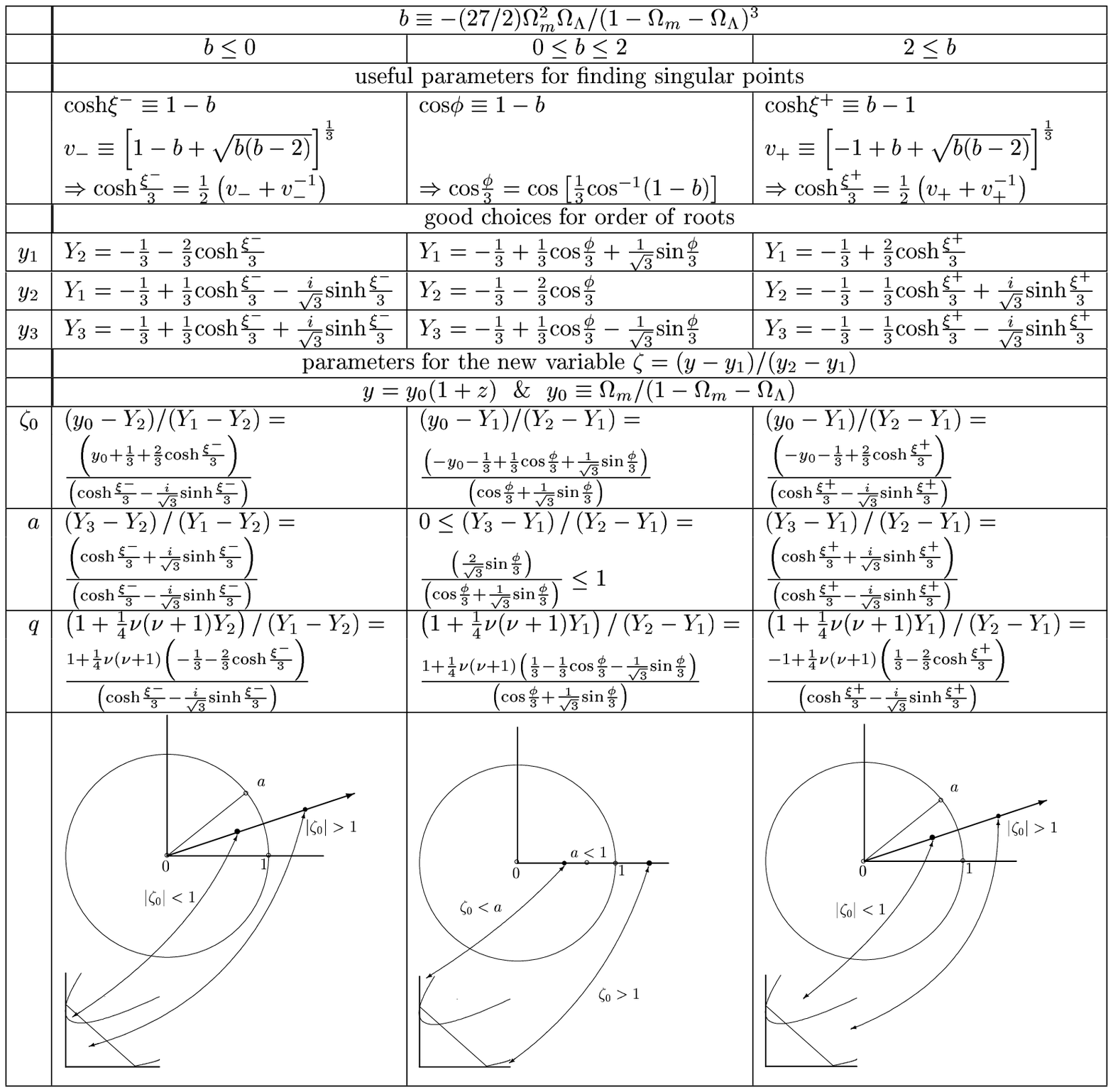]{
This table/figure contains various parameters needed for evaluating the luminosity distance
$D_{\ell}(\OM,\OL,\nu;z)$ as given by (\ref{Dell0}) and (\ref{Dellinfty}).
All parameters are functions of the cosmic parameters $\OM$, $\OL$, and the clumping parameter $\nu$.
Three columns are given for the three domains of the parameter 
$b$ separated by contours in Figure 2. 
The figure in each respective column gives the locations of the 
regular singular point $a$ in the complex $\zeta$-plane, the orbit
of $\zeta$ in (\ref{zeta0}) as a function of redshift $z$, including the starting 
point $\zeta_0$. Domains in the $\Omega_m$-$\OL$ plane that correspond to 
$|\zeta_0|<1$ and $|\zeta_0|>1$ are shown in the small reproductions of 
Figure 2 included in each column.
\label{fig3_8}}

\figcaption[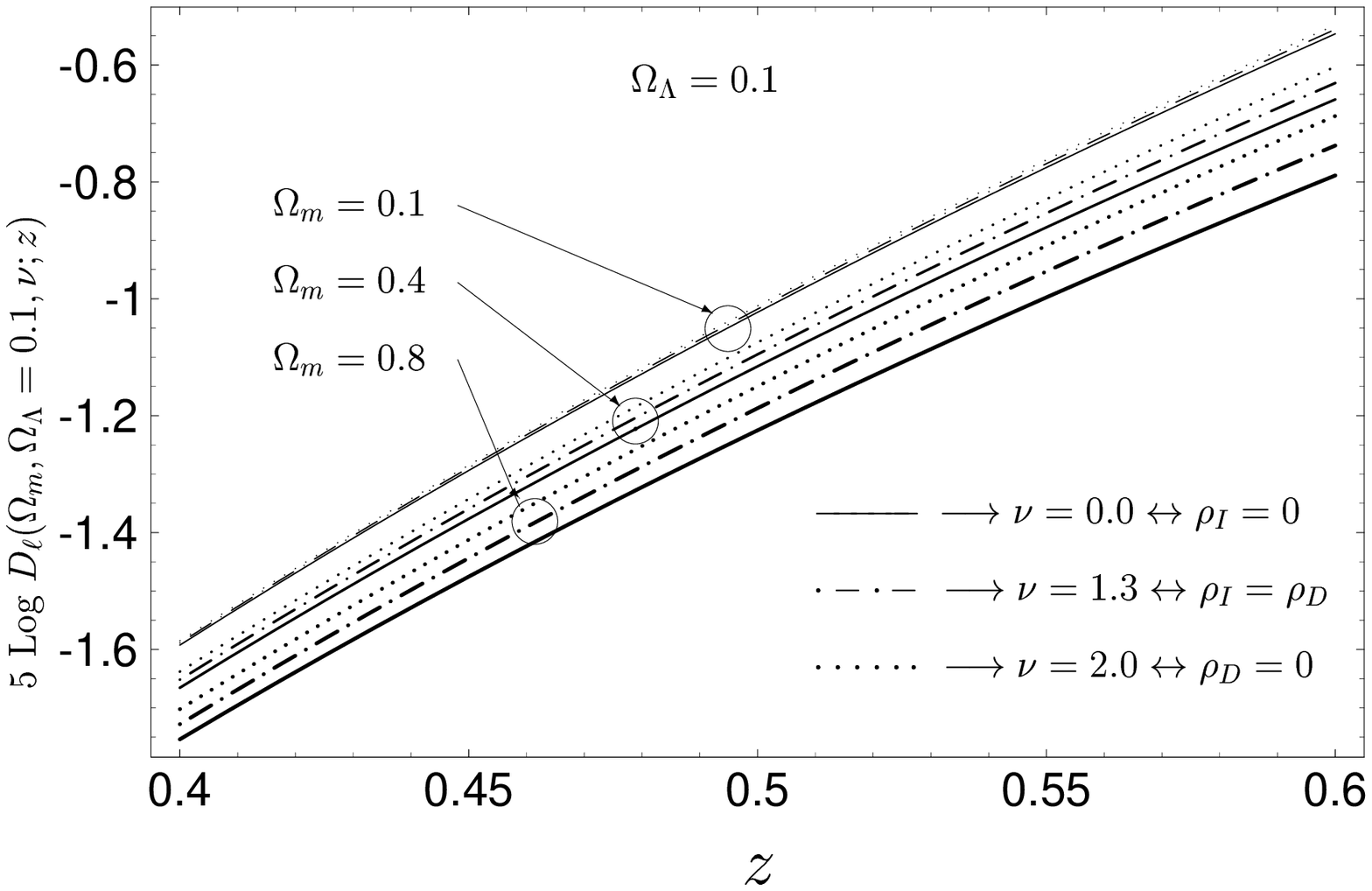]{Magnitude-redshift relation, 
$5\log_{10}{H_0\over c}D_{\ell}(\OM,\OL=0.1,\nu;z)$, 
as a function of redshift $z$ for three values of $\OM$ and 
three values of $\nu$.
\label{fig4_8}}

\figcaption[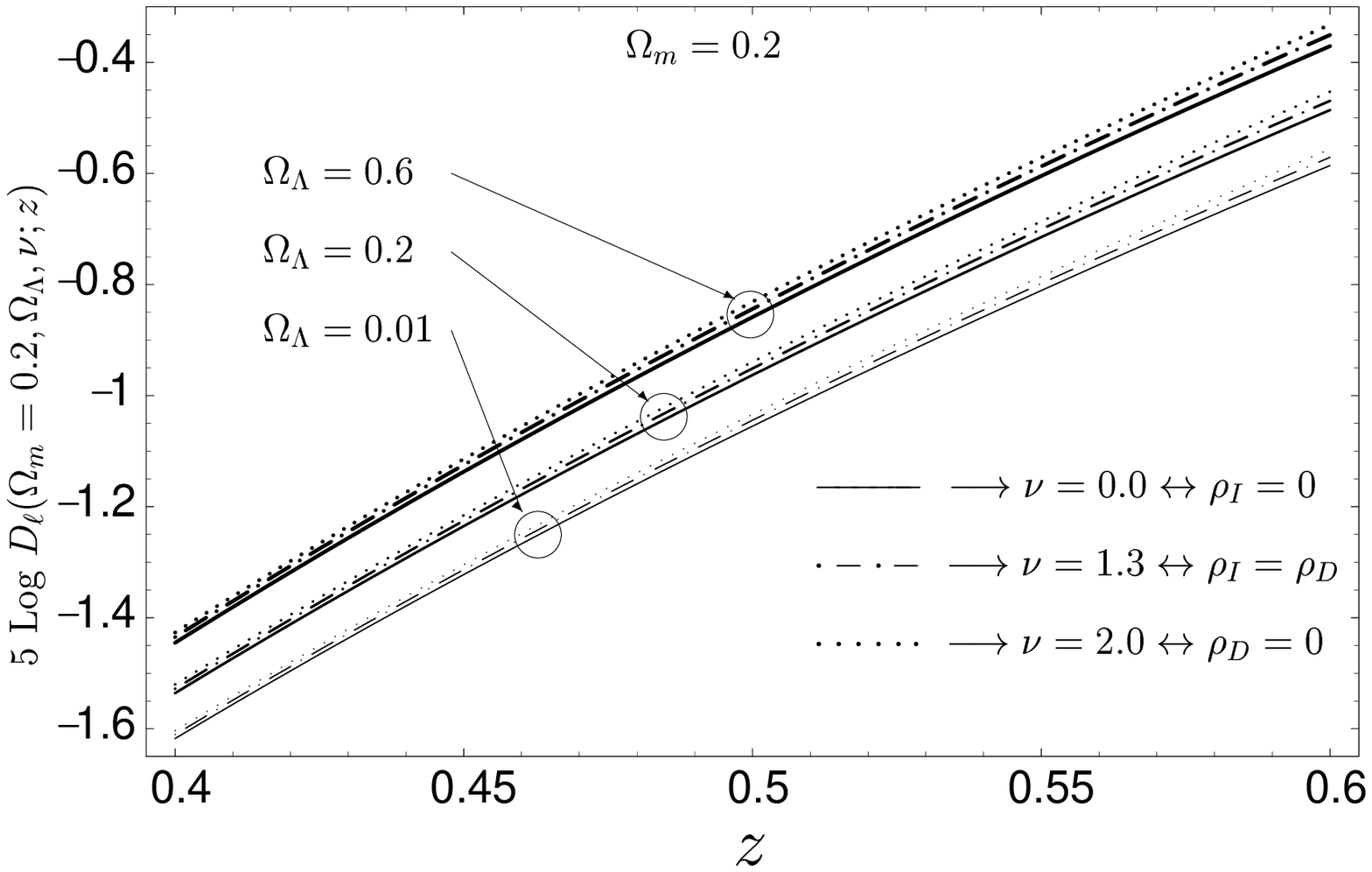]{Magnitude-redshift relation,
$5\log_{10}{H_0\over c}D_{\ell}(\OM=0.2,\OL,\nu;z)$, 
as a function of redshift $z$ for three values of $\OL$ and 
three values of $\nu$.
\label{fig5_8}}

\figcaption[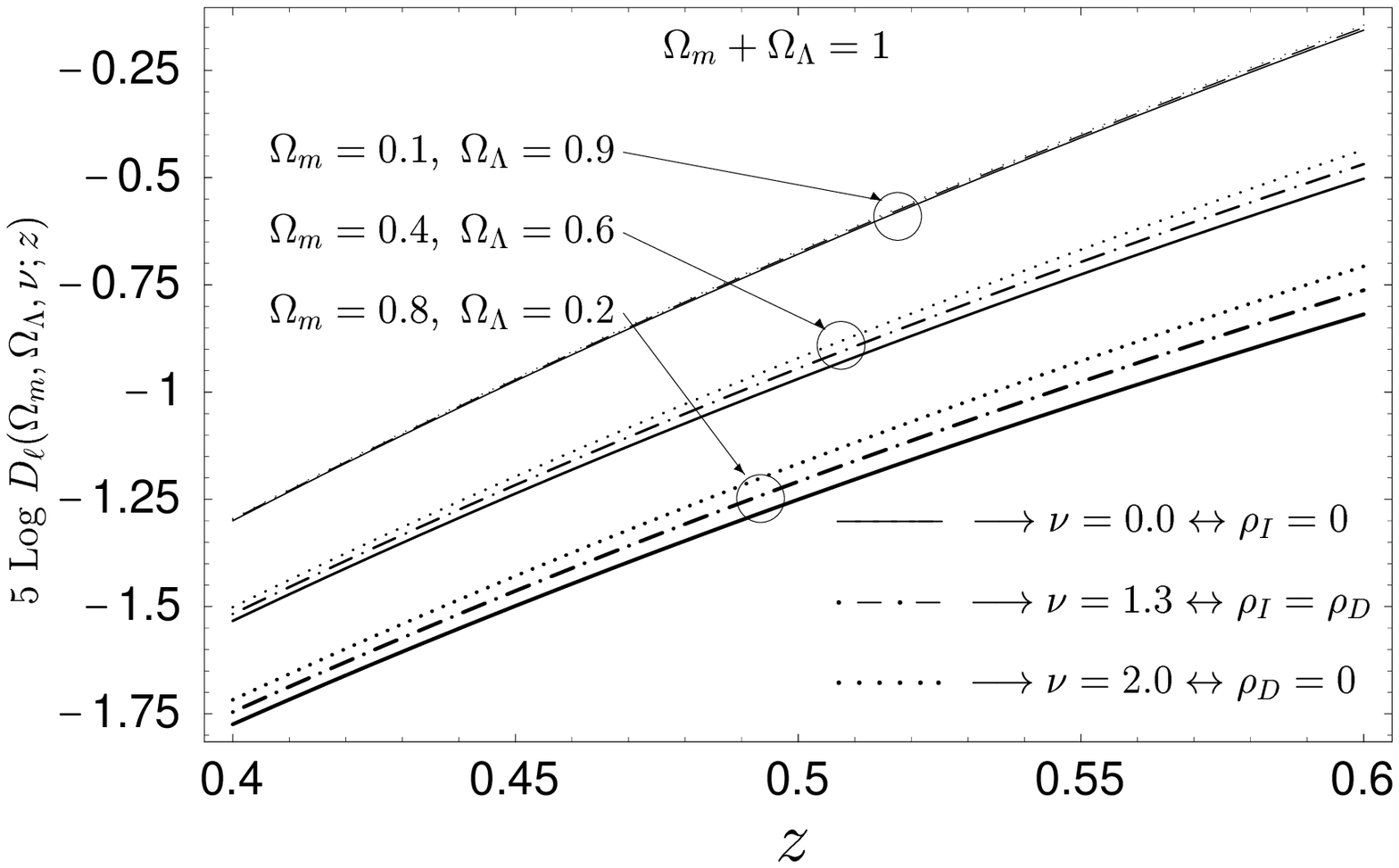]{Magnitude-redshift relation, 
$5\log_{10}{H_0\over c}D_{\ell}(\OM,\OL=1-\OM,\nu;z)$, 
as a function of redshift $z$ for three values of 
$\OM$ and $\OL$ ($\OO\equiv \OM+\OL=1$) and three values 
of $\nu$.
\label{fig6_8}}

\figcaption[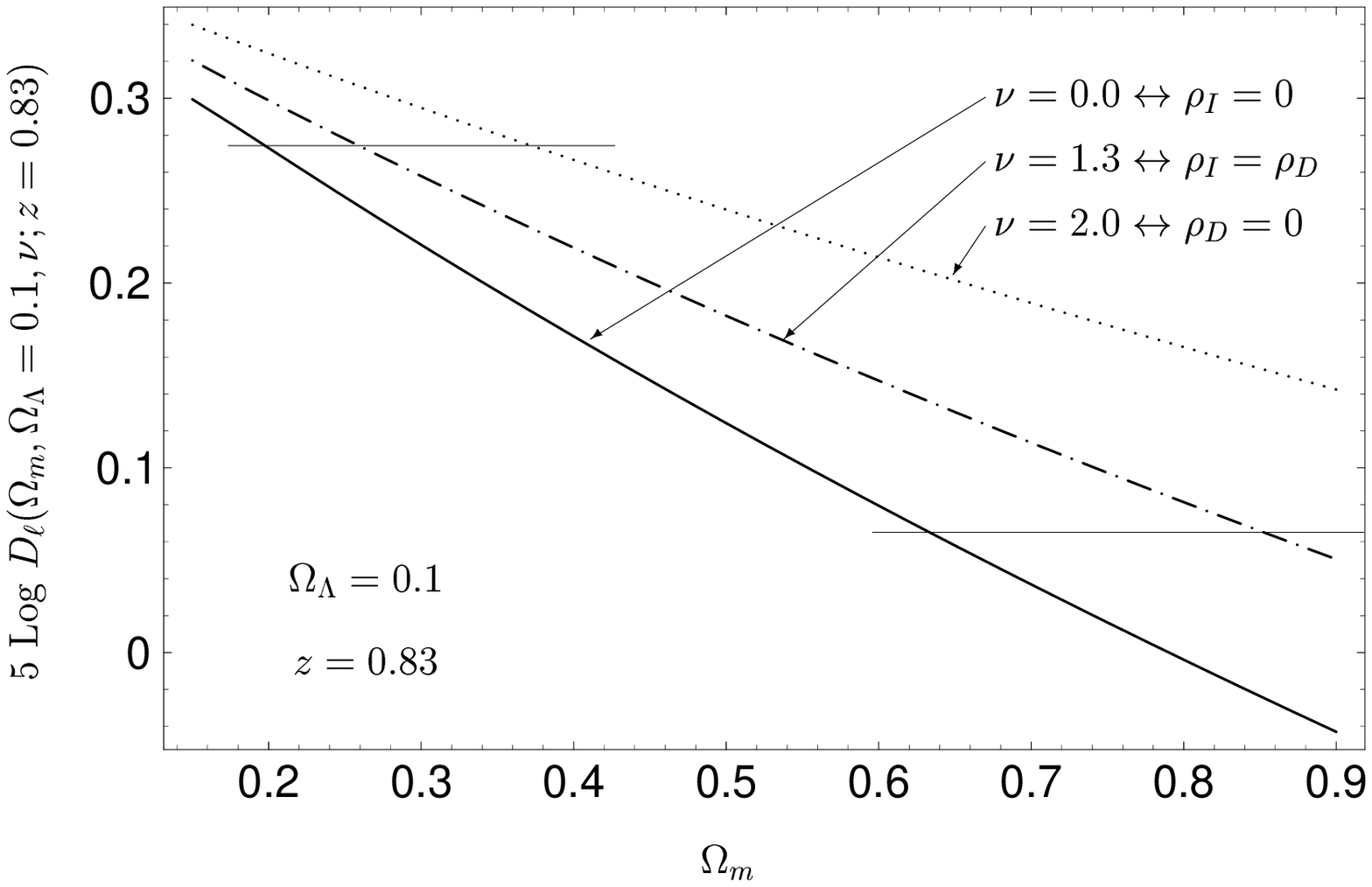]{Magnitude, $5\log_{10}{H_0\over c}D_{\ell}(\OM,\OL=0.1,\nu;z=0.83)$,
\,as a function of $\OM$ for 
three values of $\nu$.
\label{fig7_8}}

\figcaption[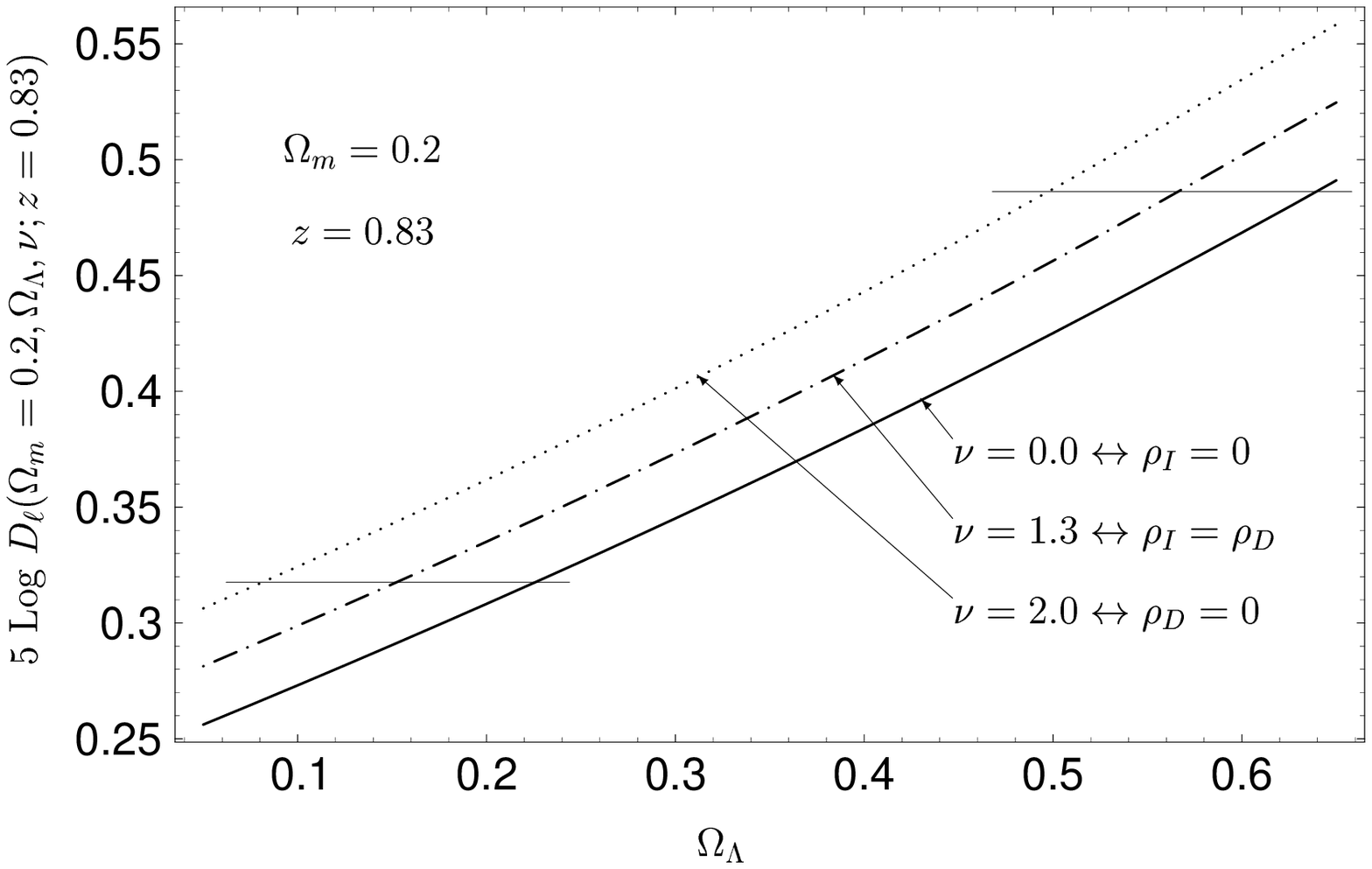]{Magnitude, $5\log_{10}{H_0\over c}D_{\ell}(\OM=0.2,\OL,\nu;z=0.83)$,
\,as a function of $\OL$ for 
three values of $\nu$.
\label{fig8_8}}

\clearpage
\plotone{fig1_8.eps}
\clearpage
\plotone{fig2_8.eps}
\clearpage
\plotone{fig3_8.eps}
\clearpage
\plotone{fig4_8.eps}
\clearpage
\plotone{fig5_8.eps}
\clearpage
\plotone{fig6_8.eps}
\clearpage
\plotone{fig7_8.eps}
\clearpage
\plotone{fig8_8.eps}


\begin{thebibliography}{}


\bibitem[Asada 1996]{AH}  Asada, H. 1997, \apj, 485, 460

\bibitem[Bertotti 1966]{BB}  Bertotti, B. 1966, Proc. Roy. Soc. London, A, 294, 195

\bibitem[Bourassa \& Kantowski 1975]{BR}  Bourassa, R.  R., \& Kantowski, R.  1975, \apj, 195, 13

\bibitem[Branch  1998]{BD}  Branch, D.  1998, \araa, 36, in press

\bibitem[Cooke \& Kantowski 1975]{CJ}  Cooke, J.  H., \& Kantowski, R.  1975, \apjl, 195, 
L11

\bibitem[Dashevskii \& Slysh 1966]{DV}  Dashevskii, V.  M., \& Slysh, V.  I.  1966,
\sovast--AJ, 9, 671

\bibitem[Dashevskii \& Zel'dovich 1965]{DVZ}  Dashevskii, V.  M., \&  Zel'dovich, Ya. B. 1965,
\sovast--AJ, 8, 854

\bibitem[Dyer \& Roeder 1972]{DC1}  Dyer, C.  C., \& Roeder, R.  C.  1972, \apjl, 174, L115

\bibitem[Dyer \& Roeder 1973]{DC2}  Dyer, C.  C., \& Roeder, R.  C.  1973, \apjl, 180, L31

\bibitem[Dyer \& Roeder 1974]{DC3}  Dyer, C.  C., \& Roeder, R.  C.  1974, \apj, 189, 167

\bibitem[Erd\'elyi 1955]{EA}  Erd\'elyi, A.  1955, Higher Transcendental Functions,
Vol.  III (New York:  McGraw-Hill)

\bibitem[Frieman 1997]{FJ}  Frieman, J.  1997, Comments Astrophys., 18, 323

\bibitem[Garnavich et al. 1998]{GP}  Garnavich, P. M. et al. 1998, \apjl, 493, L53 

\bibitem[Heun 1889]{HK}  Heun , K. 1889, Mathematische Annalen, 33, 161

\bibitem[Holz \& Wald 1998]{HD}  Holz, D. E., \& Wald, R. M. 1998, \prd, in press; astro-ph/9708036

\bibitem[Kantowski 1969]{KR}  Kantowski, R.  1969, \apj, 155, 89 

\bibitem[Kantowski 1998]{KR2}  Kantowski, R.  1998, astro-ph/9804249 

\bibitem[Kantowski et al.  1995]{KVB}  Kantowski, R., Vaughan, T., \& Branch, D.  1995,
\apj, 447, 35

\bibitem[Kayser et al. 1997]{KRa}  Kayser, R., Helbig, P., \& Schramm, T. 1997,  \aap, 318, 680

\bibitem[Mattig 1958]{MW}  Mattig, W.  1958, Astro.  Nach.  284, 109

\bibitem[Misner et al.  1973]{MTW}  Misner, C.  W., Thorne, K.  S., \& Wheeler, J.  A.
1973,  Gravitation (San Francisco:  W.  H.  Freeman)

\bibitem[Premadi et al. 1998]{PP}  Premadi, P., Martel, H., \& Matzner, R. 1998, \apj, 493, 10

\bibitem[Perlmutter et al.  1997]{PS1}  Perlmutter, S.  et al.  1997, \apj, 483, 565

\bibitem[Perlmutter et al.  1998]{PS2}  Perlmutter, S., et al.  1998, \nat, 391, 51

\bibitem[Rauch 1991]{RK}  Rauch, K.  P.  1991, \apj, 374, 83

\bibitem[Ronveaux 1995]{RA}  Ronveaux, A.  1995, Heun's Differential equations 
(Oxford:  Oxford University)

\bibitem[Sachs 1961]{SR}  Sachs, R.  K.  1961, Proc. Roy. Soc. London, A, 264, 309

\bibitem[Schneider et al.  1992]{SEF}  Schneider, P., Ehlers, J., \& Falco, E.  E.
1992, Gravitational Lenses (Berlin:  Springer-Verlag)

\bibitem[Seitz \& Schneider 1994]{SS}  Seitz, S., \& Schneider, P.  1994, \aap, 287, 349

\bibitem[Sivia  1996]{SD}  Sivia, D. S. 1996, 
Data Analysis, A Bayesian Tutorial (Oxford:  Oxford University)

\bibitem[Suto \& Matsubara 1996]{SY}  Suto, Y., \& Matsubara, T. 1996, astro-ph/9607102

\bibitem[Wambsganss et al.  1997]{WJ}  Wambsganss, J., Cen, R., Xu, G., \& Ostriker, J.
P.  1997, \apjl, 475, L81

\bibitem[Whittaker \& Watson 1927]{WE}  Whittaker, E.  T., \& Watson, G.  N.  1927,
 A Course in Modern Analysis (Cambridge:  Cambridge University Press) 

\bibitem[Zel'dovich 1964]{Zel}  Zel'dovich, Ya. B. 1964, \sovast--AJ, 8, 13
\end{thebibliography}
\end{document}